\documentclass[a4paper,fleqn,usenatbib]{mnras}

\usepackage{txfonts}

\usepackage[T1]{fontenc}
\usepackage{ae,aecompl}

\usepackage{graphicx}	
\usepackage{longtable}

\title[Non-radial modes in RR Lyrae stars]{The census of non-radial pulsation in first-overtone RR Lyrae stars of the OGLE Galactic bulge collection}

\author[H. Netzel \& R. Smolec]{
H. Netzel,$^{1}$\thanks{E-mail: henia@netzel.pl}
R. Smolec $^{1}$
\\
$^{1}$Nicolaus Copernicus Astronomical Centre, Polish Academy of Sciences, Bartycka 18, PL-00-716 Warsaw, Poland
}

\date{Accepted XXX. Received YYY; in original form ZZZ}

\pubyear{2017}

\begin{document}
\label{firstpage}
\pagerange{\pageref{firstpage}--\pageref{lastpage}}
\maketitle

\begin{abstract}
We analyzed photometry for the up-to-date collection of the first-overtone RR Lyrae stars (RRc; 11415 stars) and double-mode RR Lyrae stars (RRd; 148 stars) towards the Galactic bulge from the Optical Gravitational Lensing Experiment. We analyzed frequency spectra of these stars in search for additional, low-amplitude signals, beyond the radial modes. We focused on stars from two groups: $RR_{0.61}$ and $RR_{0.68}$. In the first group, additional low-amplitude signals have periods shorter than the first-overtone period; period ratios fall in the 0.60--0.64 range. In the second group, additional low-amplitude signals have periods longer than the first-overtone period; period ratios tightly cluster around 0.68. Altogether we have detected 960 and 147 RR Lyrae stars that belong to $RR_{0.61}$ and $RR_{0.68}$ groups, respectively, which yield the incidence rates of 8.3 and 1.3 per cent of the considered sample. We discuss statistical properties of RR Lyrae stars with additional periodicities. For $RR_{0.61}$ group we provide strong arguments that additional periodicities are connected to non-radial pulsation modes of degrees $\ell=8$ and $\ell=9$, as proposed by Dziembowski. We have also detected two double-periodic variables, with two close periodicities, similar to RR Lyrae variable V37 in NGC 6362. Properties of these peculiar variables, which may form a new group of double-mode pulsators, are discussed.
\end{abstract}




\section{Introduction}
Till recently, RR Lyrae stars were considered to be a textbook example of simple classical pulsating stars. Significant majority pulsate in one radial mode. In majority of stars it is the radial fundamental mode (RRab stars). In a smaller number of stars, it is the radial first-overtone mode (RRc stars). We also know double-mode pulsation in the fundamental mode and the first overtone mode (RRd stars) or in the fundamental mode and the second overtone. The latter group is not very numerous, only a few members are known \citep[e.g.][]{benko1, benko2}. Period of pulsations for RR Lyrae stars is in the 0.2--1 d range. For a long time, the only flaw on this simple picture was the Blazhko effect, which is a quasi-periodic amplitude and/or phase modulation of pulsations observed in a subgroup of RR Lyrae stars. It was discovered more than a hundred years ago (Blazhko 1907). The Blazhko effect is a frequent phenomenon among RRab stars, for which the incidence rate of modulated stars can be as high as 50 per cent \citep{jurcsik_bl, benko2, ngc6362}. The modulation is less frequent in RRc stars. The analysis of the OGLE Galactic bulge RRc stars led to the incidence rate of 5.6 per cent \citep{moje_bl}. The Blazhko effect was also detected in RRd stars \citep{jurcsik_bl, smolec_rrd_bl, soszrrd}.

Analysis of space photometry revealed that in many RR Lyrae stars additional small-amplitude signals are observed, which cannot be due to radial modes. The first new group that seems to appear at unexpected position in the Petersen diagram (i.e. diagram of period ratio versus the longer period) consists of stars in which the radial first-overtone mode dominates (so either RRc or RRd stars). The additional shorter-period and low-amplitude signal forms period ratio around $0.60-0.64$. We denote this group $RR_{0.61}$. This period ratio suggests that the additional mode cannot correspond to radial pulsations. After first discovery of this additional signal in the RRd star AQ Leo \citep{aqleo}, it was found in many more RRc and RRd stars from ground-based photometry \citep[see e.g.][and references therein]{olech, moje_o4_061, jurcsik_rrc} and from space-based photometry \citep[see e.g.][and references therein]{068kepler, 061_k2, kurtz2016}.
An explanation of the $RR_{0.61}$ group was proposed by \cite{dziembowski}. In this scenario, additional signal forming period ratio $0.60-0.64$ is a harmonic of the non-radial mode of moderate degree ($\ell=8,9$).

Another interesting group are RRc stars with the additional long-period signal forming period ratio $P_{\rm 1O}/P_x$ around 0.68 with the first-overtone period. This $RR_{0.68}$ group was reported for the first time by \cite{moje_o4_068}. Observed period ratio indicates, that the additional signals have period longer than the period of the fundamental mode, which was never detected in these stars. The problem with an explanation of this group was discussed by \cite{dziembowski}.

\cite{zdenek} reported discovery of another previously unknown group of double-mode RR Lyrae stars with the additional short-period signal of relatively large amplitude, forming period ratios from the range $0.68-0.72$. Explanation of the nature of this additional signal is still missing.

The Optical Gravitational Lensing Experiment \citep[OGLE,][]{ogle} is a large-scale sky variability survey. OGLE regularly monitors the brightness of stars from the Magellanic System, the Galactic bulge and the Galactic disk in {\it I} and {\it V} filters. Since 2010 the fourth phase \citep[OGLE-IV,][]{o4} is ongoing. The main advantages of the OGLE photometry are a long time base, high sampling and very good quality of photometry. Despite the fact, that OGLE is a ground-based project, the long time base together with a numerous sample make OGLE data the perfect source to search for rare phenomena in pulsating stars and to study long-term behavior of pulsations. In the Galactic bulge fields the OGLE collection of variable stars counts over 38~000 RR Lyrae stars \citep{o4bulge, sosz_morerrl}. Among them over 27~480 are RRab stars, 11~415 are RRc stars, and 148 are RRd stars. 

We present the analysis of the full sample of first-overtone RR Lyrae stars from the Galactic bulge. Main aim of this work is to provide the complete census of $RR_{0.61}$ and $RR_{0.68}$ stars. Method of the analysis is discussed in Sec.~\ref{sec.data}. In Sec.~\ref{sec.results} we present an overview of the results, while in Sec.~\ref{sec.061} and \ref{sec.068} $RR_{0.61}$ and $RR_{0.68}$ stars are discussed in detail. In Sec.~\ref{sec.jak_v37} we discuss two interesting objects detected additionally during this analysis. Our findings are summarized in Sec.~\ref{sec.summary}.

\section{Data and analysis}\label{sec.data}
The data used for analysis were gathered during the fourth phase of the OGLE project, which started in 2010 and is ongoing \citep{o4}. Eight seasons of observations carried out in 2010-2017 in the Galactic bulge are available for the analysis. We used observations in the {\it I} filter only, as these are more numerous than in the {\it V} filter. The input sample for the analysis comprises 11~415 RRc stars. Because of the numerous sample, the study was largely automatic.


To search for additional periodicities, automatic script, based on Fourier transform and consecutive pre-whitening method, was used. The first step was to fit the dominant frequency and its harmonics to the data. Frequencies detected in the Fourier transform were fitted to the data in the form:
\begin{equation}
m(t)=A_0+\sum A_k\sin(2\pi f_{k}t+\phi_k)
\label{Eq.fit}
\end{equation}
where $f_k$, $A_k$ and $\phi_k$ are frequencies, amplitudes and phases of peaks detected in the power spectrum. 

Only frequencies which fulfilled the criterion $A_k/\sigma(A_k)>4$ were included in the fit, where $A_k$ is an amplitude of the signal and $\sigma(A_k)$ is its error. 

In the data we often detect strong signals related to trends and non-stationary pulsation, which increase the noise level in the Fourier transform and hamper the detection of additional periodicities. Trend in the power spectrum manifests as signal or power excess in the low frequency range. High fraction of RRc stars have non-stationary dominant periods, which manifest in power spectrum as unresolved power at the position of the dominant frequency after prewhitening. These signals usually result from slow, often irregular pulsation period changes. In order to remove these signals we used a form of time-dependent prewhitening method, which was proposed by \cite{068kepler}. For the application of the time-dependent prewhitening to the OGLE data see \cite{moje_o3}. In most cases, this method was successful in removing these unwanted signals. After applying time-dependent prewhitening with the first-overtone and its harmonics, the automatic procedure proceeded with the search for additional signals.

The search was performed in the following way: the power spectrum was calculated from 0 to 40 cycles per day. The procedure determined the highest signal in this range. If the signal had a signal-to-noise ratio, S/N, higher than 4, then it was added to the fit. Otherwise, the analysis was finished. When a new signal was detected, the procedure decided whether it is independent, or it corresponds to a linear frequency combination of the already detected signals. In the latter case, no new independent frequency was introduced to Eq.~\ref{Eq.fit}, but appropriate linear combination term. The procedure also excluded signals located very close to integer frequencies, i.e. $f=1,2,3$ c\,d$^{-1}$ as these are most likely of diurnal origin.

If no new signals were detected in the power spectrum, or the number of independent frequencies reached 16, then the analysis was finished. The last step was to remove outliers with the $4\sigma$ criterion.

The output of the procedure were frequency lists for all RRc stars. Then we picked those RRc stars, in which the procedure found signals characteristic for interesting groups: $RR_{0.61}$ and $RR_{0.68}$. For $RR_{0.61}$ group we selected those stars, for which period ratio of the additional signal with the first overtone falls into $0.6-0.64$ range. In the case of $RR_{0.68}$ stars, criterion for range of period ratios was $0.66-0.7$. In case of some stars, time-dependent prewhitening did not remove remnant power in the position of trend or the first overtone. Aliases of remaining signals can be misinterpreted as additional signals. Therefore, the last step of the analysis was visual inspection of prewhitened power spectra in frequency range from 0 to 10 c\,d$^{-1}$ for selected stars. Dubious cases were analyzed manually. 

We detected additional signals in other stars as well, however, they do not form any new group of multi-mode RR Lyrae stars. Moreover, we do not observe combination frequencies in case of these additional signals, which suggests, that they may be due to contamination.

Additionally, we manually analyzed 148 RRd stars from the Galactic bulge.

\section{Results}\label{sec.results}
Analysis of the RRc sample resulted in a discovery of 949 $RR_{0.61}$ stars and 147 $RR_{0.68}$ stars. Corresponding incidence rates are 8.3 per cent and 1.3 per cent. Analysis of RRd stars resulted in the detection of 11 $RR_{0.61}$ stars and no $RR_{0.68}$ stars. The corresponding incidence rate for RRd $RR_{0.61}$ stars is 7 per cent.
List of $RR_{0.61}$ stars with their properties is in Tab.~A1, sample of which is presented in Tab.~\ref{tab:results061} for a reference. Some stars have more than one row in Tab.~A1. In these stars we detected more than one signal forming period ratio in the $0.60-0.64$ range. These stars will be discussed in more detail in Sec.~\ref{Subsec.kilkaciagow}. List of $RR_{0.68}$ with their properties is in Tab.~A2, sample of which is presented in Tab.~\ref{tab:results068}. In both tables we provide name of a star, its first-overtone period, period of the additional signal, ratio between two periods, amplitude of the first overtone and ratio of amplitude of the additional signal to the amplitude of the first overtone mode. Last column contains remarks on individual stars.

\begin{figure}
\centering
\includegraphics[width=0.5\textwidth]{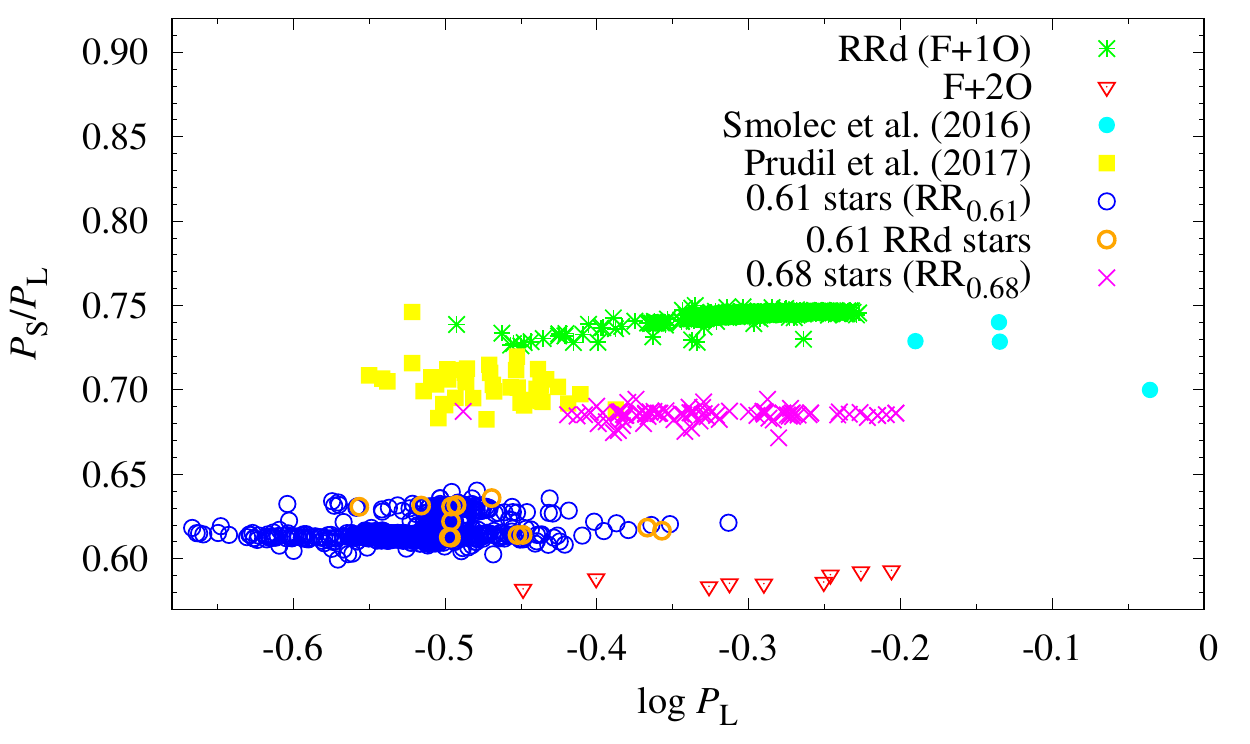}
\caption{Petersen diagram for multi-mode pulsations in RR Lyrae stars. $RR_{0.61}$ group consists of stars selected during this study and stars reported previously by other studies.}
\label{Fig.ogle_results}
\end{figure}

$RR_{0.61}$ and $RR_{0.68}$ stars are plotted in the Petersen diagram of multi-mode RR Lyrae stars in Fig.~\ref{Fig.ogle_results}. $RR_{0.61}$ stars are plotted with blue open circles and RRd stars are plotted with orange open circles. $RR_{0.68}$ stars are plotted with pink crosses. In Sec.~\ref{sec.061} and \ref{sec.068} we discuss $RR_{0.61}$ and $RR_{0.68}$ stars in detail.
We also detected two interesting stars with additional modes, which are similar to V37 from the NGC 6362 analyzed by \cite{ngc6362}. They are discussed in Sec.~\ref{sec.jak_v37}.

\begin{table*}
\begin{minipage}{\textwidth}
\centering
\caption{Sample table with properties of $RR_{0.61}$ stars. Full table is in the Appendix. Consecutive columns provide name of a star, its first-overtone period, period of the additional signal, ratio between the two periods, amplitude of the first overtone and ratio of amplitude of the additional signal to the amplitude of the first overtone mode. Last column contains remarks on individual stars: `comb.' - there is a combination between the first overtone and the additional signal, `bl' - Blazhko effect is present in the star, `cand.' - additional signal is weak (around $S/N=4$) or its daily aliases are higher, $0.5f_x$ - subharmonic of the additonal signal is detected.}
\label{tab:results061}
\begin{tabular}{lcccccc}
ID & $P_{\rm 1O}$\thinspace [d] &  $P_{\rm X}$\thinspace [d] & $P_{\rm X}/P_{\rm 1O}$  & $A_{\rm 1O}$\thinspace [mag]   &  $A_{\rm X}/A_{\rm 1O}$   & Remarks \\
\hline
OGLE-BLG-RRLYR-00710 & 0.28749643(7) & 0.180509(1) & 0.62787 & 0.1420(7) & 0.021 &  \\ 
OGLE-BLG-RRLYR-00815 & 0.3117463(3) & 0.192034(1) & 0.61599 & 0.120(2) & 0.041 &  \\ 
OGLE-BLG-RRLYR-01014 & 0.30330353(6) & 0.186214(1) & 0.61395 & 0.1226(4) & 0.022 &  \\ 
OGLE-BLG-RRLYR-01067 & 0.28294809(2) & 0.1736623(8) & 0.61376 & 0.1426(2) & 0.013 &  \\ 
OGLE-BLG-RRLYR-01088 & 0.29471870(5) & 0.1811678(9) & 0.61471 & 0.1319(4) & 0.021 &  \\ 
OGLE-BLG-RRLYR-01097 & 0.2923084(4) & 0.179451(1) & 0.61391 & 0.1357(5) & 0.026 &  \\ 
\end{tabular}
\end{minipage}
\end{table*}

\begin{table*}
\begin{minipage}{\textwidth}
\centering
\caption{Sample table with properties of $RR_{0.68}$ stars. Full table is in the Appendix. Consecutive columns provide name of a star, its first-overtone period, period of the additional signal, ratio between the two periods, amplitude of the first overtone and ratio of amplitude of the additional signal to the amplitude of the first overtone mode. Last column contains remarks on individual stars: `comb.' - there is a combination between the first overtone and the additional signal, `bl' - Blazhko effect is present in the star.}
\label{tab:results068}
\begin{tabular}{lcccccc}
ID & $P_{\rm 1O}$\thinspace [d] &  $P_{\rm X}$\thinspace [d] & $P_{\rm 1O}/P_{\rm X}$  & $A_{\rm 1O}$\thinspace [mag]   &  $A_{\rm X}/A_{\rm 1O}$   & Remarks \\
\hline
OGLE-BLG-RRLYR-01064 & 0.31486155(4) & 0.458607(6) & 0.68656 & 0.1239(3) & 0.016 &  \\ 
OGLE-BLG-RRLYR-01152 & 0.34759513(10) & 0.507535(4) & 0.68487 & 0.1084(5) & 0.04 & comb. \\ 
OGLE-BLG-RRLYR-01408 & 0.3853534(3) & 0.56176(1) & 0.68598 & 0.112(1) & 0.04 &  \\ 
OGLE-BLG-RRLYR-01436 & 0.3242810(5) & 0.47942(2) & 0.6764 & 0.120(3) & 0.04 &  \\ 
OGLE-BLG-RRLYR-01808 & 0.3161546(2) & 0.460225(5) & 0.68696 & 0.0937(8) & 0.063 &  \\ 
OGLE-BLG-RRLYR-03053 & 0.2898407(1) & 0.423478(6) & 0.68443 & 0.1273(9) & 0.042 & bl \\ 
\end{tabular}
\end{minipage}
\end{table*}

\section{$RR_{0.61}$ stars}\label{sec.061}
During this analysis of RRc and RRd stars we detected altogether 960 (949 RRc and 11 RRd) stars falling into the $RR_{0.61}$ group, where 744 stars are new detections (734 RRc and 10 RRd). Combining with previous results from the analysis of the OGLE-III data \citep{moje_o3} and the analysis of two fields from the OGLE-IV data \citep{moje_o4_061}, the total number of $RR_{0.61}$ stars from the OGLE data is 994 RRc and 12 RRd stars. 

\begin{figure}
\centering
\includegraphics[width=0.5\textwidth]{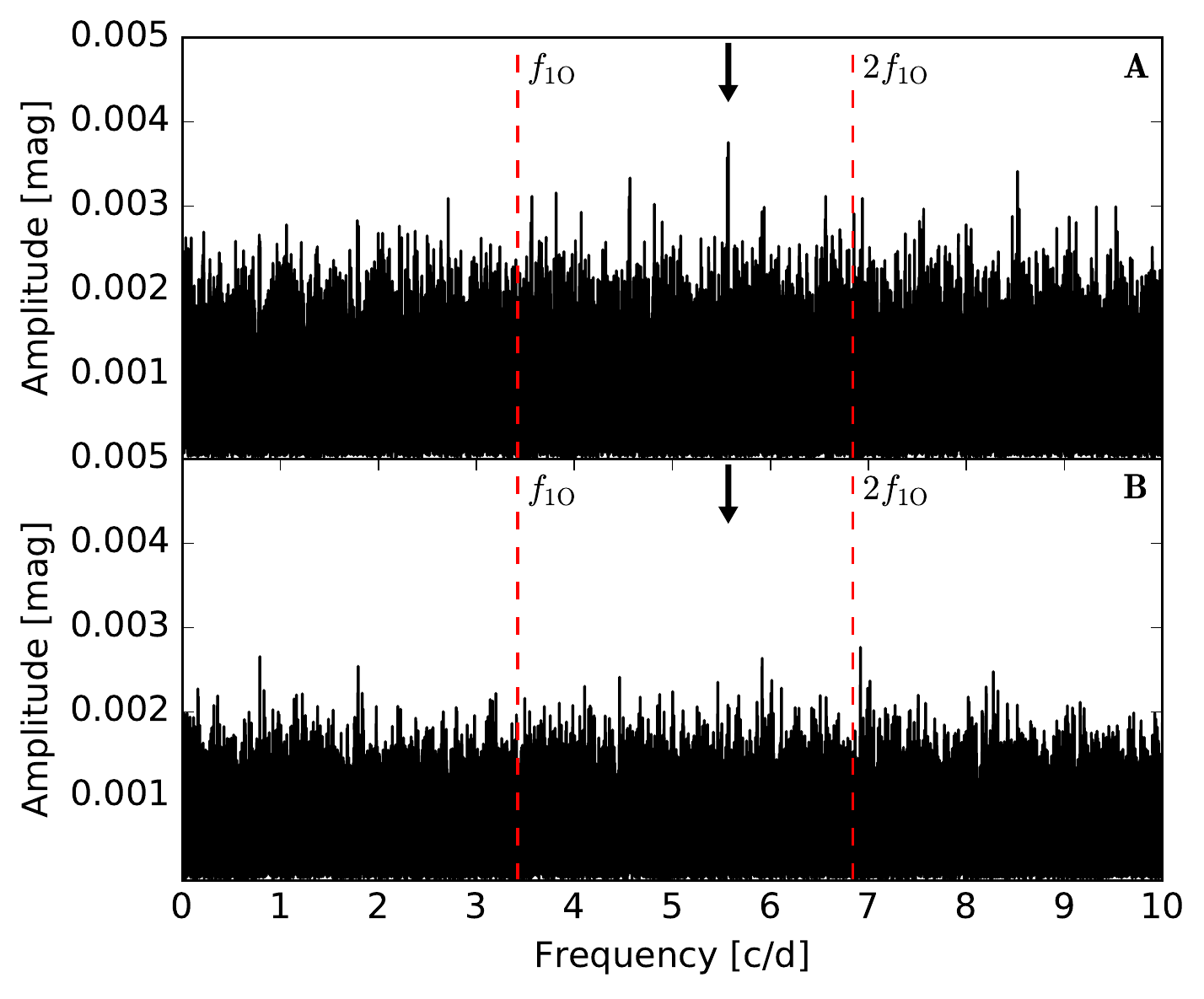}
\caption{Example of a star, which was detected as $RR_{0.61}$ when O-III data were used (panel A), but was not detected as $RR_{0.61}$ in OGLE-IV data (panel B).}
\label{Fig.02077}
\end{figure}

Several RRc stars detected in the previous studies were not found during this analysis. An example of such star, OGLE-BLG-RRLYR-02077, is plotted in Fig.~\ref{Fig.02077}. This star was classified as $RR_{0.61}$ based on the analysis of the OGLE-III data, power spectrum of which is presented in the panel A of Fig.~\ref{Fig.02077}. Position of the additional signal is marked with an arrow. In the panel B of Fig.~\ref{Fig.02077} we plotted power spectrum using OGLE-IV data from this analysis. Despite the lower noise level, there is no signal at the position of an arrow. This fact may be a consequence of the variable nature of the additional signal. \cite{moje_o4_061} presented in their fig.~10. analysis of the $RR_{0.61}$ star on the season-by-season basis. Due to variability of the amplitude of the signal, the same star would have different classification depending on the season of observations. \cite{068kepler} in their fig.~7 presented variability in time of the amplitudes of the additional signal and its subharmonics in four $RR_{0.61}$ stars observed by {\it Kepler}.

\begin{figure}
\centering
\includegraphics[width=0.5\textwidth]{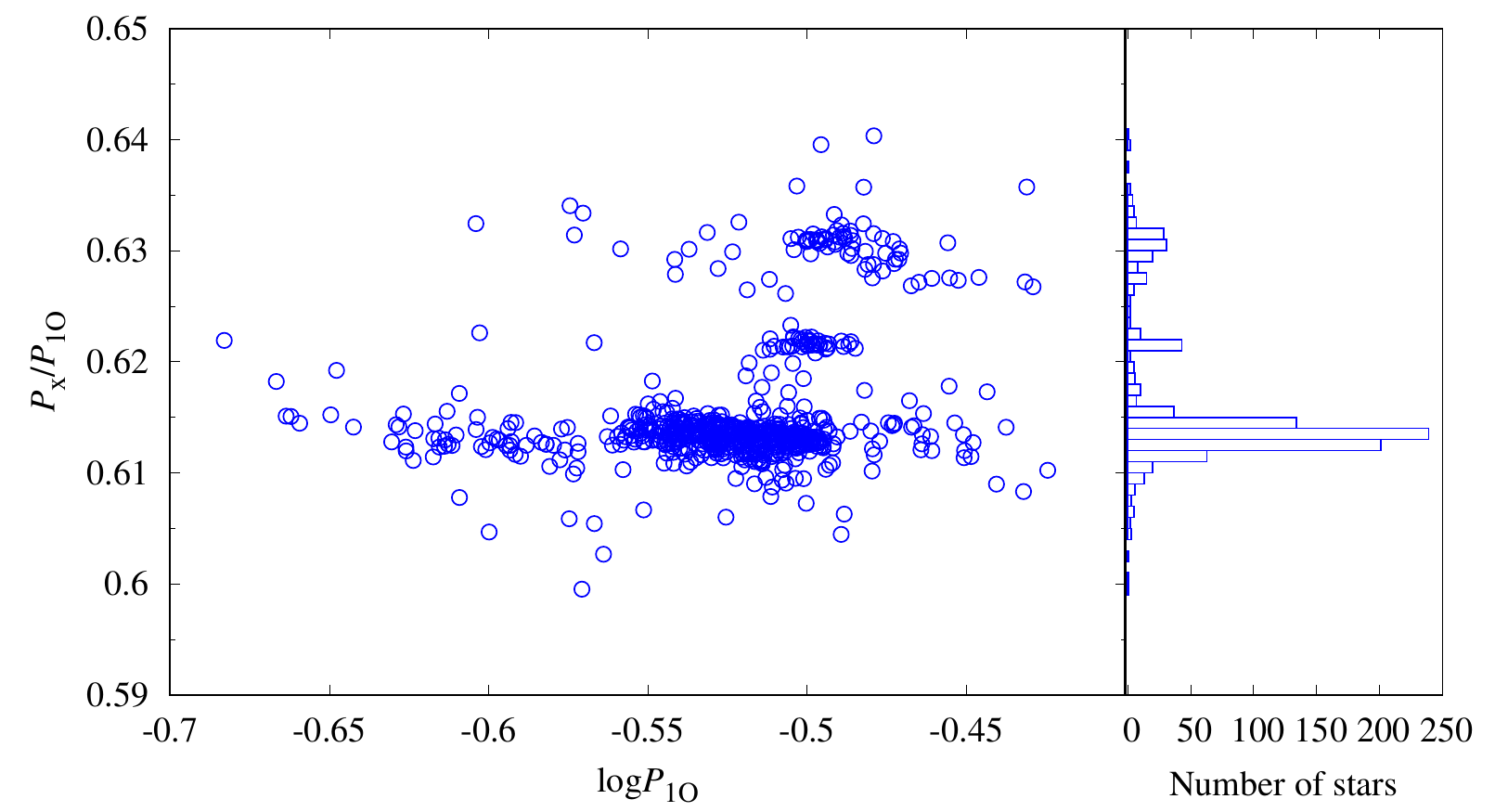}
\caption{Petersen diagram for RRc and RRd stars with additional 0.61 signal detected in this study. Right panel shows the distribution of period ratios.}
\label{Fig.061pet}
\end{figure}

Petersen diagram for $RR_{0.61}$ stars identified in this study is presented in Fig.~\ref{Fig.061pet}. On the right panel we provided distribution of period ratios. Three sequences are clearly visible in the diagram. Based on minima in the distribution of period ratios we define the borderlines between the sequences. Stars were classified as members of the bottom sequence when period ratio was below 0.62. For period ratios between 0.62 and 0.625 stars were classified as members of the middle sequence and for period ratios above 0.625 as members of the top sequence. Clearly, the lowest sequence is the most populated and is centered at period ratio 0.613. The middle sequence is very tight and centered at period ratio 0.622 and the top sequence is centered at period ratio 0.631 with larger scatter than in the middle sequence. By numbers, the bottom, middle and top sequences count 792, 112 and 221 stars, respectively. These numbers do not add up to 960. This is due to the fact, that in some stars we detected signals corresponding to more than one sequence.

\begin{figure}
\centering
\includegraphics[width=0.5\textwidth]{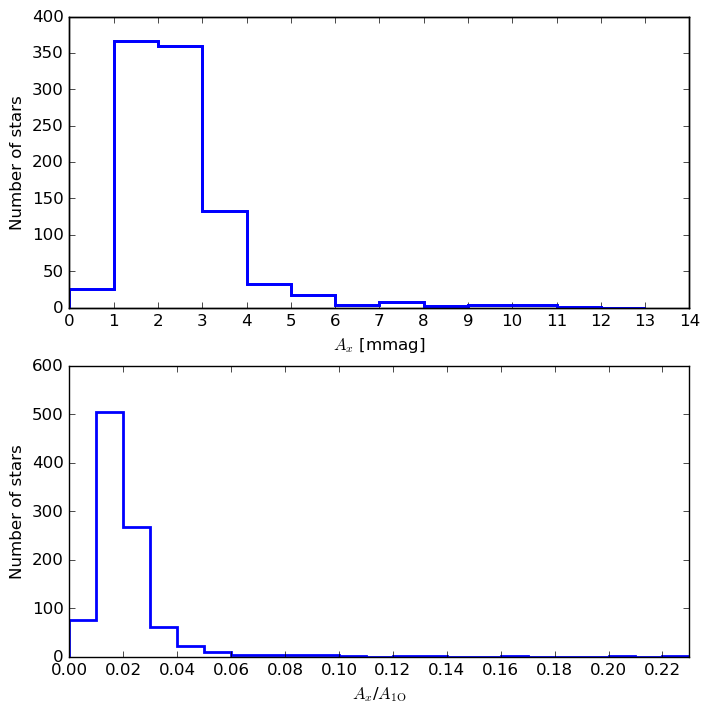}
\caption{In the top panel we show distribution of amplitudes of the additional signal in $RR_{0.61}$ stars. In the bottom panel we plotted amplitude ratio between the additional signal and the first overtone. One signal for each star included.}
\label{Fig.061amp}
\end{figure}

\begin{figure}
\centering
\includegraphics[width=0.5\textwidth]{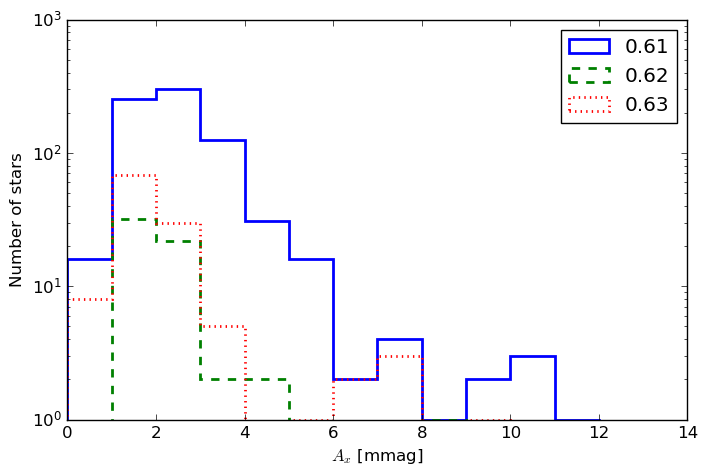}
\caption{Distribution of amplitudes in $RR_{061}$ stars with distinction to three sequences. One signal for each star included.}
\label{Fig.061amp3ciagi}
\end{figure}

Amplitudes of the additional signals are low, in the mmag range. For the majority of stars, the amplitude is $1-4$ mmag, as visible in Fig.~\ref{Fig.061amp} which shows the distribution of amplitudes of additional signals for all $RR_{0.61}$ stars. The highest amplitude is 11 mmag, and the lowest is 0.4 mmag. In the bottom panel of Fig.~\ref{Fig.061amp} we show the amplitude ratio between the additional mode and the first overtone. For majority of stars this ratio is below 5 per cent, and the peak of the distribution is at 2 per cent. In Fig.~\ref{Fig.061amp3ciagi} we plotted the distribution of amplitudes for the additional signal with the distinction for signals corresponding to the three sequences. Stars with the highest amplitudes are almost only from the lowest sequence. The middle and top sequences correspond to low amplitudes, mostly below 4 mmag. 

\subsection{Subharmonics}
In some stars observed by space telescopes, we observe subharmonics of the additional signal centered at the position of $0.5\thinspace f_{x}$ and/or $1.5\thinspace f_{x}$ \citep{068kepler}. From ground-based data we see almost exclusively signals at $0.5\thinspace f_{x}$. Just as signals detected at $f_x$, these signals have complex structure. Typically we observe a power excess centered at $0.5f_x$, rather than a single and coherent peak. According to \cite{dziembowski}, signals at subharmonic frequencies, $0.5f_x$, are non-radial modes of moderate degrees, $\ell=8$, and $\ell=9$, and the signals at $f_x$ are harmonics. Due to cancellation effect, it is more difficult to detect the actual non-radial mode, but it is easier to identify its harmonic. Therefore, in the ground-based data, we detect mostly the harmonics, that form the period ratios around 0.6-0.64 with the radial first overtone. We detected signals at $0.5\thinspace f_{x}$ for 114 stars from our sample (111 RRc and 3 RRd).

\begin{figure}
\centering
\includegraphics[width=0.5\textwidth]{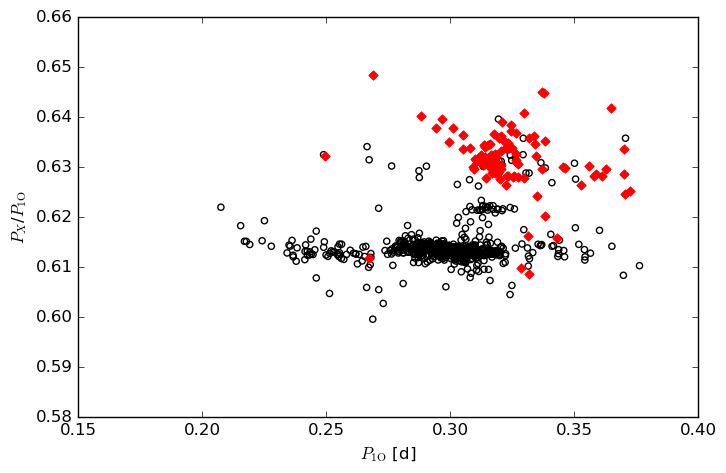}
\caption{Petersen diagram for OGLE $RR_{0.61}$ stars. Stars without subharmonic in power spectra are plotted with open black circles. For stars in which the signal at the subharmonic frequency is detected, corresponding signal is plotted with red diamonds.}
\label{Fig.subh_pet}
\end{figure}

In the model proposed by \cite{dziembowski}, the top sequence in the Petersen diagram arises due to harmonics of $\ell=8$ non-radial modes, the bottom sequence arises due to harmonics of $\ell=9$ non-radial modes, and the middle sequence arises due to linear combination frequency of these two modes.
Due to differences in cancellation effect for modes with different $\ell$, it should be easier to detect signal corresponding to the non-radial mode with $\ell=8$ than with $\ell=9$. In Fig.~\ref{Fig.subh_pet} we plotted the Petersen diagram for $RR_{0.61}$ stars from our sample. Stars in which we did not detect signal at $0.5f_x$, i.e. the non-radial mode, are plotted with black open circles. Stars in which we detected this signal are plotted with red diamonds.
All stars with subharmonic, except eight, correspond to the top sequence, i.e. indeed the non-radial mode of $\ell=8$ is more easily detected than non-radial mode of $\ell=9$. This finding strongly supports the Dziembowski model.

\begin{figure}
\centering
\includegraphics[width=0.5\textwidth]{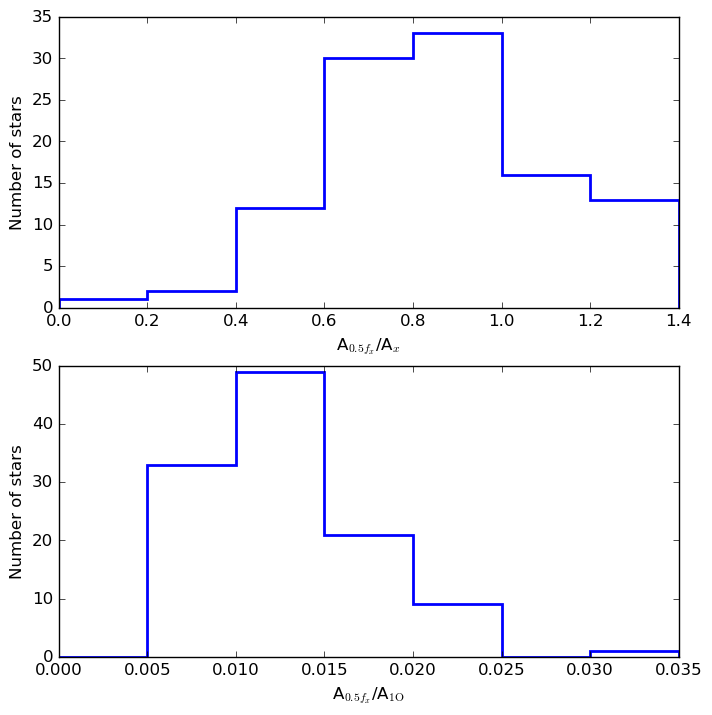}
\caption{In the top panel we show distribution of amplitude ratio of subharmonic signal to the additional signal. In the bottom panel we plotted amplitude ratio of subharmonic signal to the first overtone.}
\label{Fig.subh_hist}
\end{figure}

In Fig.~\ref{Fig.subh_hist} we plotted histograms of amplitude ratio between the signal at $0.5f_x$ and the signal at $f_x$ (top panel), and between the signal at $0.5f_x$ and the first-overtone mode (bottom panel). Amplitude of the $0.5f_x$ signal constitutes mostly 0.5 to 2 per cent of the amplitude of the first overtone and 60 to 100 per cent of the signal at $f_x$. In 35 out of 114 stars we detected higher amplitude the $0.5f_x$ signal than of the $f_x$ signal.


In seven stars, which are not members of the $RR_{0.61}$ group, we detected additional signal, which forms period ratio with the first overtone, $P_{\rm 1O}/P_y$ around 0.79. This additional signal has period longer than the first-overtone period. However, period corresponding to its harmonic, which is not detected in the power spectrum, forms period ratio 0.60-0.64 with the first overtone. The fact, that in 31 per cent of $RR_{0.61}$ stars in which both signals are detected (35 out of 114) we observe higher amplitude of the signal at $0.5f_x$ than of the signal at $f_x$ (top panel of Fig.~\ref{Fig.subh_hist}) suggests, that these seven stars are in fact $RR_{0.61}$ stars in which we observe only the $0.5f_x$ signal. Properties of these stars are collected in Tab.~\ref{tab:tylkosubharm}. In the fourth column of Tab.~\ref{tab:tylkosubharm} we provide the ratio of the period corresponding to the harmonic of the detected signal to the first-overtone period. In six out of seven stars this period ratio corresponds to the top sequence, for which it is the most likely to directly detect non-radial mode, i.e. signal corresponding to $0.5f_x$. 

\begin{table*}
\begin{minipage}{\textwidth}
\centering
\caption{Table with properties of stars in which signal fitting into the position of $0.5 f_{\rm x}$ was detected. Consecutive columns provide ID of a star, period of the first overtone ($P_{\rm 1O}$), and of the detected signal ($P_{0.5 f_{\rm x}}$), period ratio of the harmonic of the detected signal with the first-overtone period ($P_{0.5 f_{\rm x}}/\left( 2P_{\rm 1O}\right)$), amplitude of the first overtone and amplitude ratio.}
\label{tab:tylkosubharm}
\begin{tabular}{lccccc}
ID & $P_{\rm 1O}$\thinspace [d] &  $P_{0.5 f_{\rm x}}$\thinspace [d] & $P_{0.5 f_{\rm x}}/\left( 2P_{\rm 1O}\right)$  & $A_{\rm 1O}$\thinspace [mag]   &  $A_{0.5 f_{\rm x}}/A_{\rm 1O}$    \\
\hline
OGLE-BLG-RRLYR-02815 & 0.31485827(8) & 0.396265(7) & 0.62928 & 0.1265(5) & 0.017   \\ 
OGLE-BLG-RRLYR-04622 & 0.3220768(2) & 0.407628(7) & 0.63281 & 0.1358(9) & 0.034  \\ 
OGLE-BLG-RRLYR-05746 & 0.32689535(6) & 0.412324(6) & 0.63067 & 0.1284(3) & 0.015  \\ 
OGLE-BLG-RRLYR-06501 & 0.32539511(7) & 0.415580(6) & 0.63858 & 0.1160(4) & 0.014   \\ 
OGLE-BLG-RRLYR-09926 & 0.38240378(3) & 0.471320(4) & 0.61626 & 0.1129(2) & 0.014 \\
OGLE-BLG-RRLYR-31948 & 0.29931521(4) & 0.381558(3) & 0.63739 & 0.1277(3) & 0.012  \\
OGLE-BLG-RRLYR-32466 & 0.32554603(8) & 0.411304(6) & 0.63171 & 0.1168(4) & 0.016   \\
\end{tabular}
\end{minipage}
\end{table*}

\subsection{Stars with more than one signal}\label{Subsec.kilkaciagow}
As it was reported by \cite{moje_o4_061}, some stars have more than one signal in the frequency range corresponding to $0.60-0.64$ period ratio. We detected signals corresponding to two sequences in 101 stars from our sample (all classified as RRc) and to three sequences in 34 stars from our sample (33 RRc and 1 RRd).  These 34 stars give an excellent opportunity to check whether indeed the middle sequence arises due to linear combination frequency of $\ell=8$ and $\ell=9$ non-radial modes as proposed by \cite{dziembowski}.

\begin{figure}
\centering
\includegraphics[width=0.5\textwidth]{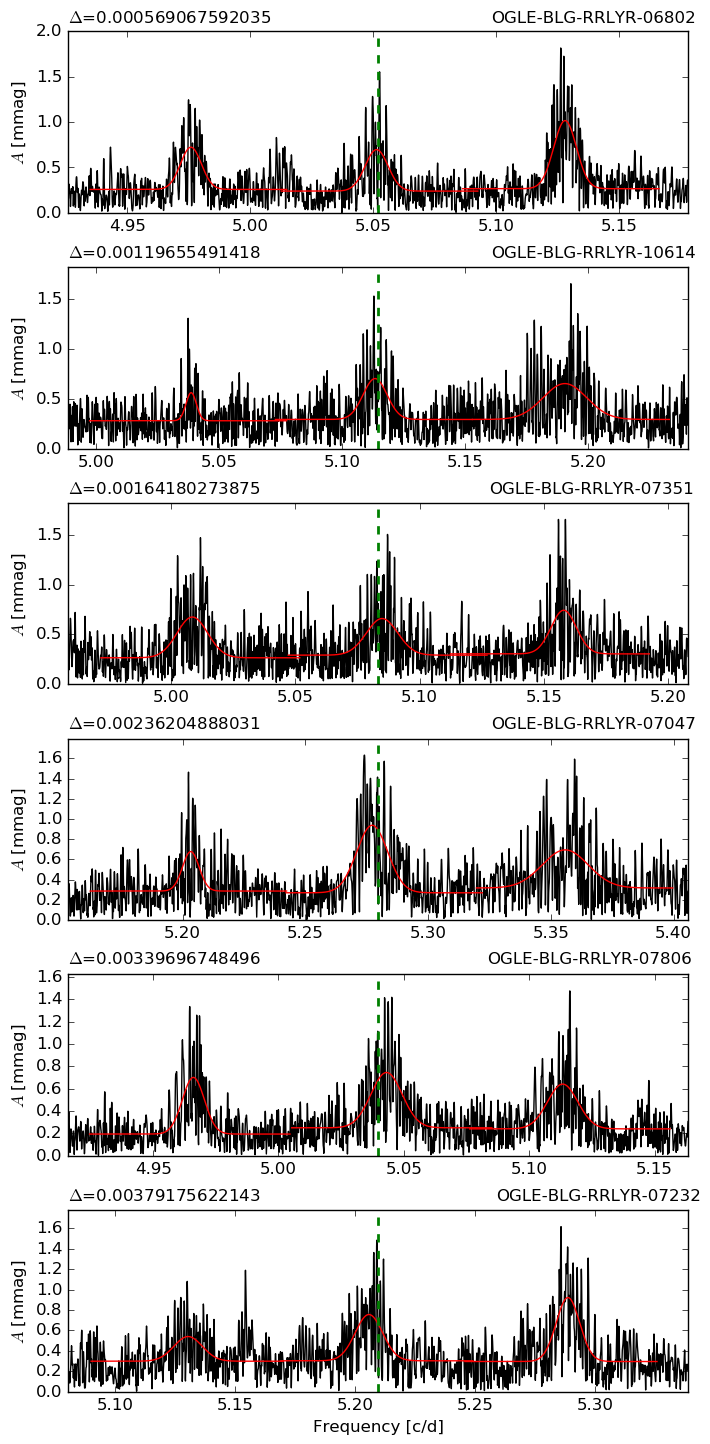}
\caption{Examples of stars with three signals in the $0.6-0.64$ frequency range. Stars are sorted with increasing $\Delta$ and each correspond to one bin from Fig.~\ref{Fig.3ciagi_diff_gauss}. Red line shows Gaussian fits to the signals. Green dashed line is placed at $0.5(f_{0.61}+f_{0.63})$.}
\label{Fig.gauss_examples}
\end{figure}

\begin{figure}
\centering
\includegraphics[width=0.5\textwidth]{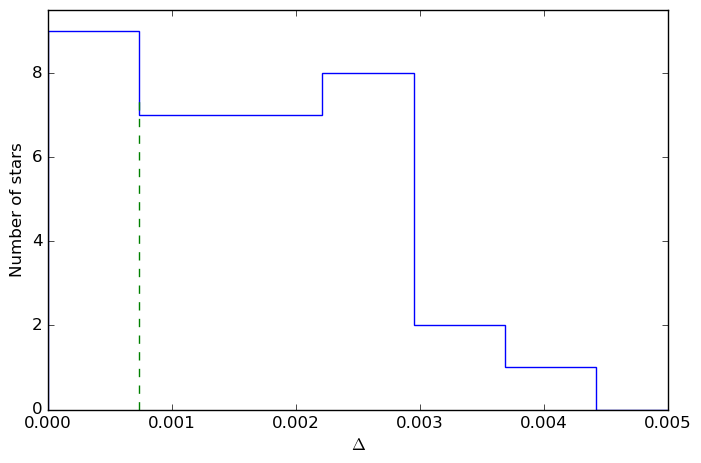}
\caption{Histogram of the deviation parameter $\Delta=|f_{\rm middle}-0.5(f_{061}+f_{063})|$ in stars with signals corresponding to three sequences. Green dashed line corresponds to resolution of the power spectrum, which is the same for majority of stars included in the figure.}
\label{Fig.3ciagi_diff_gauss}
\end{figure}

In this case, the frequency of the signal from the middle sequence ($f_{middle}$) should be an average of the frequencies of the signals from the bottom ($f_{0.61}$) and the top ($f_{0.63}$) sequences. As these signals vary in time, peaks in the power spectra are very broad (Fig.~\ref{Fig.gauss_examples}) and merely taking the frequencies of the highest peaks within the power excesses, to test the postulated relation, would lead to non reliable results. Therefore we fitted these excess powers with Gaussian function (plotted with red line in Fig.~\ref{Fig.gauss_examples}) and adopted the centroid as the frequency of the signals. Green dashed line in Fig.~\ref{Fig.gauss_examples} is placed then at $0.5(f_{0.61}+f_{0.63})$. Then we constructed the deviation parameter, which we defined as the following frequency difference, $\Delta=|f_{\rm middle}-0.5(f_{061}+f_{063})|$. In Fig.~\ref{Fig.3ciagi_diff_gauss} we show the distribution of the deviation parameter in the discussed sample. The green dashed line corresponds to the adopted resolution of the power spectrum. Its location may suggest that in several stars the discussed relation between position of the three signals is not fulfilled. This is not the case however, as these signals are wide. Frequency spectra of stars in Fig.~\ref{Fig.gauss_examples} are sorted by increasing deviation parameter and in the last panel we plotted a star with the largest deviation parameter in the considered sample. It is clear that the calculated deviations are always smaller than full width at half maximum (FWHM) of the fitted Gaussians and the discussed relation is well fulfilled given the nature of the signals.


\section{$RR_{0.68}$ stars}\label{sec.068}
\begin{figure}
\centering
\includegraphics[width=0.5\textwidth]{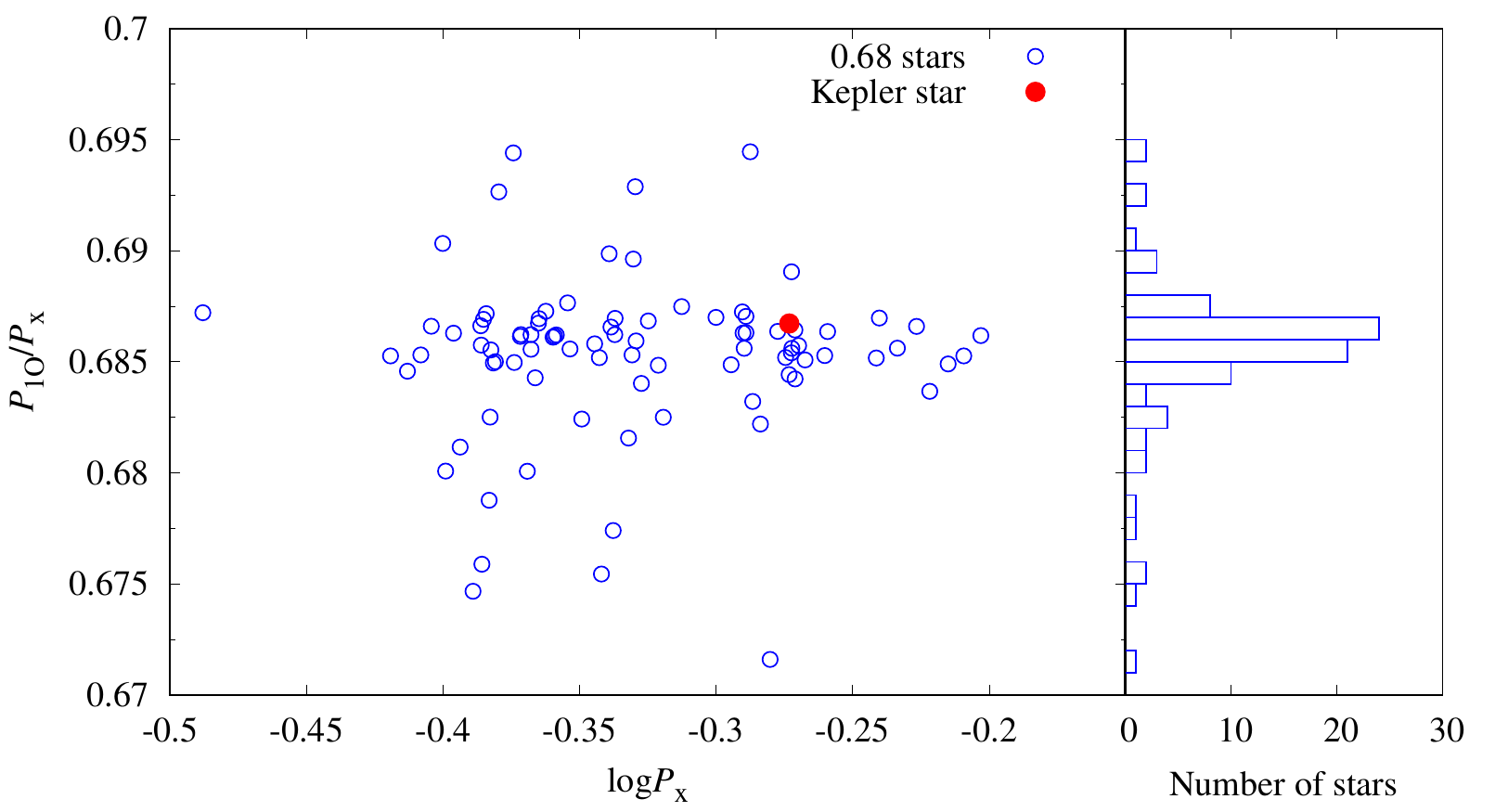}
\caption{Zoom in the Petersen diagram for $RR_{0.68}$ stars. Stars found in the OGLE data are plotted with blue open circles. Star detected in the {\it Kepler} data is plotted with red filled circle. Right panel shows distribution of period ratio among these stars.}
\label{Fig.068pet}
\end{figure}

Stars with additional long-period signals, forming period ratio around 0.68 with the first overtone mode, were detected for the first time in the OGLE data \citep{moje_o4_068}. A few more stars were reported in \cite{pta_068}. One star, KIC 9453114, is known from the {\it Kepler} data \citep{068kepler}. This star belongs to the $RR_{061}$ group and contains additional low frequency signals, one of which places this star also in the $RR_{0.68}$ group.
 During this study we identified 147 RRc stars with this additional signal. We did not detect this signal in any RRd star. Corresponding incidence rate for RRc stars is 1.3 per cent. 128 stars are new detections. In Fig.~\ref{Fig.068pet} we show zoom in the Petersen diagram for all known OGLE $RR_{0.68}$ stars together with KIC 9453114. On the right panel we show distribution of period ratios in these stars. The distribution is centered on period ratio 0.686, as previously reported by \cite{moje_o4_068}. However, there are several outliers with smaller or higher period ratios.

All $RR_{0.68}$ stars detected in this study are collected in Table A2, sample of which is presented in Tab.~\ref{tab:results068}. Consecutive columns provide ID, first-overtone period, period of the additional signal, period ratio, first-overtone amplitude, amplitude ratio of additional signal to the first-overtone mode. Remarks are given in the last column.

We note that contrary to additional signals in $RR_{0.61}$ group, additional variability in $RR_{0.68}$ stars is always coherent. Amplitudes and phases of these signals do not vary in time.

\begin{figure}
\centering
\includegraphics[width=0.5\textwidth]{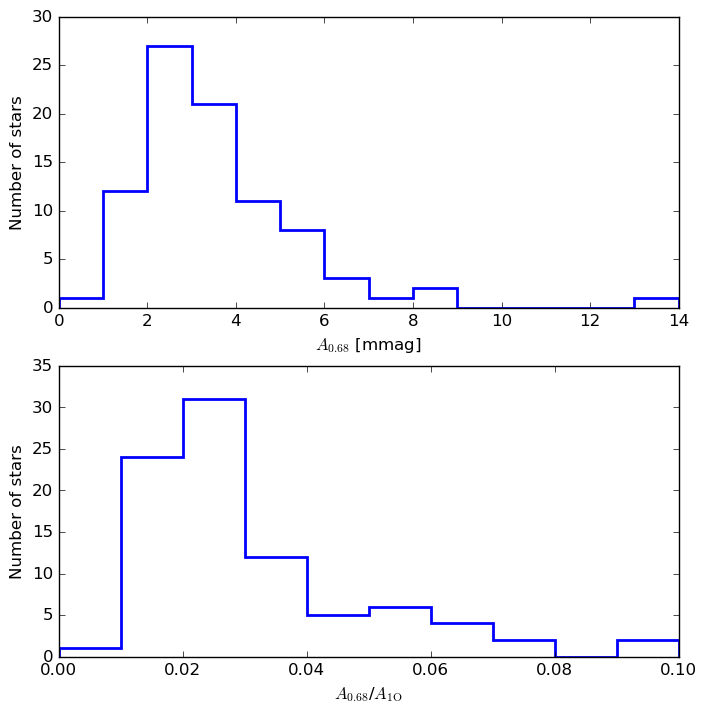}
\caption{Top panel shows distribution of amplitudes of the additional signal forming period ratio 0.68 with the first overtone. Bottom panel shows distribution of amplitude ratio between the additional signal and the first overtone mode.}
\label{Fig.068amp_hist}
\end{figure}

In Fig.~\ref{Fig.068amp_hist} we present distribution of amplitudes of the additional signal (top panel) and distribution of amplitude ratio of additional signal to the first overtone (bottom panel). Amplitude of this signal is 2-4 mmag in the majority of stars, which constitutes 1 to 3 per cent of the first overtone amplitude.

Two $RR_{0.68}$ stars are particularly interesting: OGLE-BLG-RRLYR-11167 and OGLE-BLG-RRLYR-30601. In OGLE-BLG-RRLYR-11167 we detected two signals corresponding to two period ratios: 0.68271 and 0.68106. It is the only star which has two rows in the Tab. A2. This star also has the Blazhko effect and was discussed in detail in \cite{moje_bl}. We note that separation between the signals does not correspond to the Blazhko modulation period.

In OGLE-BLG-RRLYR-30601, besides the first overtone and the signal corresponding to the 0.68 period ratio, $f_{0.68}$, we detected one more signal, $f_x$, which has period in between the period of the first overtone and the period of the $f_{0.68}$ signal. Both additional signals form combinations with the first overtone. Period ratio between the first overtone and the $f_x$ signal is 0.7695, so higher than for RRd stars. In the Petersen diagram, the two periodicities, $f_x$ and $f_{\rm 1O}$ place OGLE-BLG-RRLYR-30601 at the long-period end of the double-mode sequence of the High Amplitude Delta Scuti (HADS) stars pulsating in the fundamental mode and in the first overtone. If OGLE-BLG-RRLYR-30601 is indeed HADS, then it would show that the $f_{0.68}$ signal can be present in other stars than RR Lyrae \citep[see also][ for the possible detection in Classical Cepheids]{cep068}, and it would be the first star with this peculiar periodicity which pulsates in two radial modes simultaneously. However, classification of OGLE-BLG-RRLYR-30601 as HADS is not certain, because Fourier decomposition parameters fit very well typical values for RRc stars.

%
%
%
%

\section{OGLE-BLG-RRLYR-11754 and OGLE-BLG-RRLYR-15059}\label{sec.jak_v37}

\begin{figure}
\centering
\includegraphics[width=0.5\textwidth]{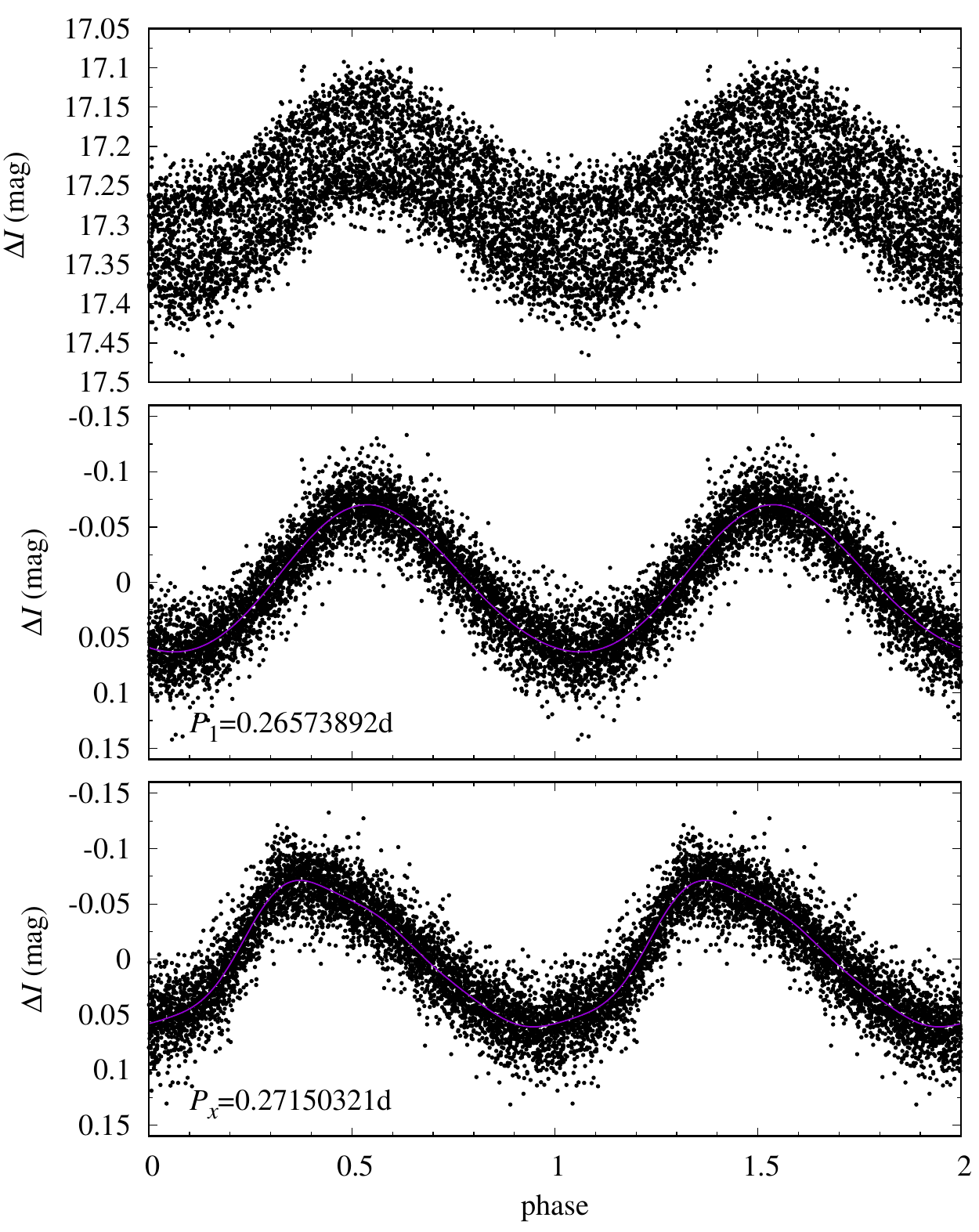}
\caption{Light curve of OGLE-BLG-RRLYR-11754 phased with the dominant periodicity ($P_1$) is plotted in the top panel. Disentangled light curves for the dominant and the additional signal are plotted in the middle and bottom panels, respectively.}
\label{Fig.11754_lc}
\end{figure}

\begin{figure}
\centering
\includegraphics[width=0.5\textwidth]{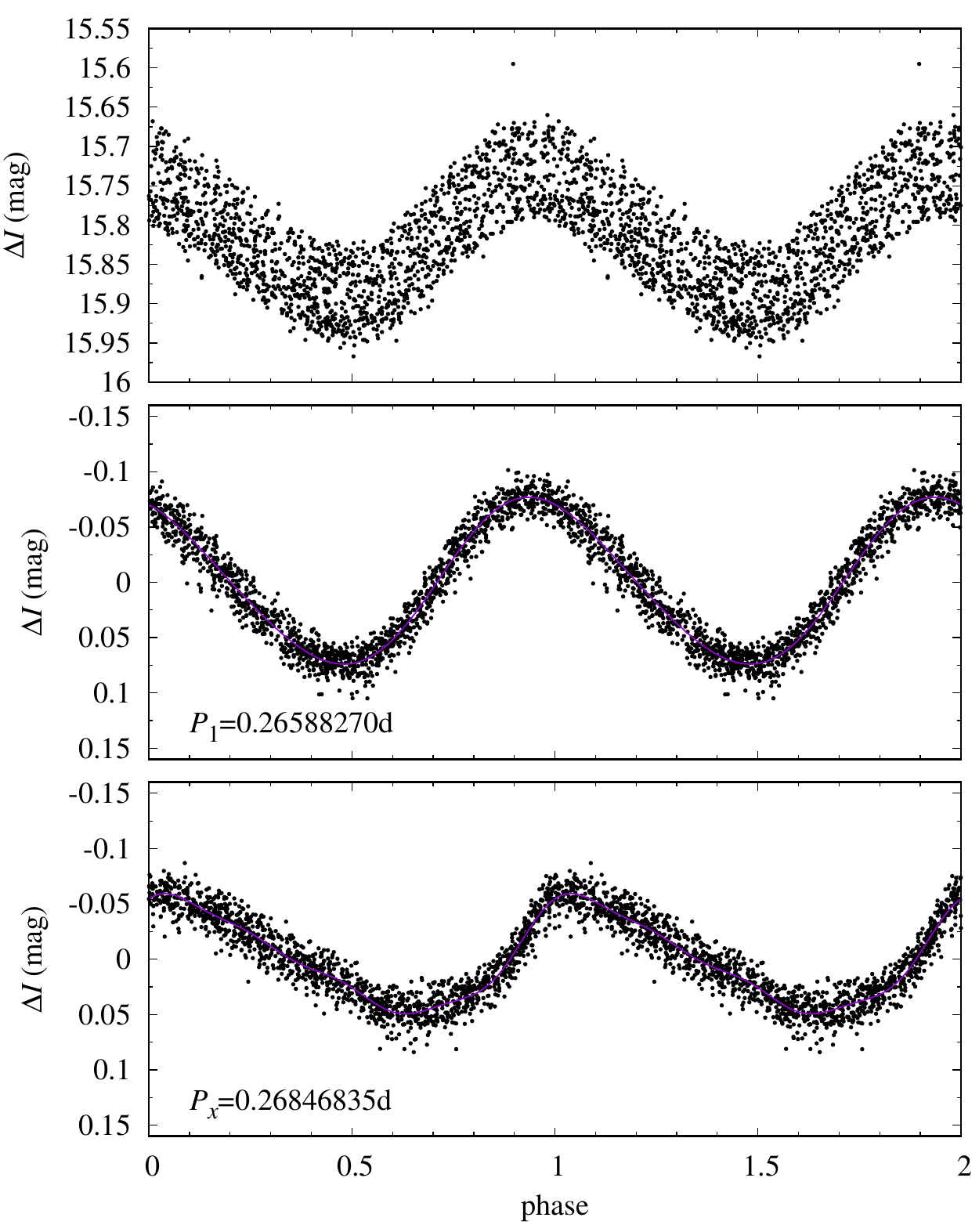}
\caption{The same as Fig.~\ref{Fig.11754_lc}, but for OGLE-BLG-RRLYR-15059.}
\label{Fig.15059_lc}
\end{figure}

\begin{figure}
\centering
\includegraphics[width=0.5\textwidth]{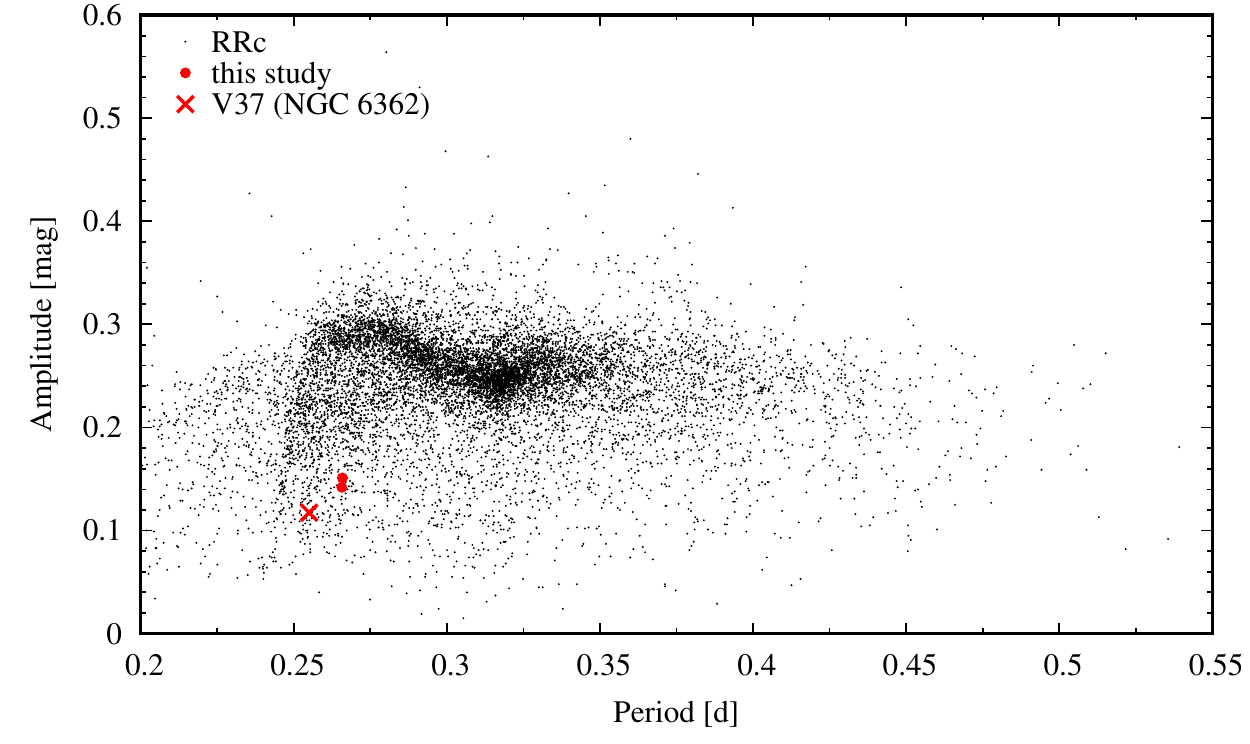}
\caption{Period -- amplitude diagram for RRc stars (black dots). OGLE-BLG-RRLYR-11754 and OGLE-BLG-RRLYR-15059 are plotted with red points. V37 star (NGC 6362) is plotted with red cross.}
\label{Fig.jakv37_amp}
\end{figure}

We detected two peculiar stars, OGLE-BLG-RRLYR-11754 and OGLE-BLG-RRLYR-15059, which are very similar to RRc star V37 in the globular cluster NGC 6362 analyzed by \cite{ngc6362}. In the period amplitude diagram for RRc stars (Fig.~\ref{Fig.jakv37_amp}) all three variables are located in the short-period and low-amplitude region. Both stars we discuss here have light curves that immediately suggest multiperiodicity or modulation (see top panels of Figs \ref{Fig.11754_lc} and \ref{Fig.15059_lc}). In Tab.~\ref{tab:11754} and Tab.~\ref{tab:15059} we collected all frequencies detected in OGLE-BLG-RRLYR-11754 and OGLE-BLG-RRLYR-15059, respectively. Consecutive columns provide frequencies of signals, their amplitudes and phases. Last two columns provide two possible interpretations of the detected signals: modulation or beating. In power spectra of both stars we detected the additional signal with the frequency very close to the frequency of the first overtone. In the modulation scenario this additional signal corresponds to the modulation sidepeak, and its separation with the first-overtone frequency corresponds to the modulation frequency, $\nu_m$. In the beating scenario this additional signal is treated as an independent frequency and is denoted in both tables as $\nu_x$. As in V37, modulation scenario is unlikely. In the power spectra of both stars we observe very incomplete high-order multiplet components on the lower frequency side of the harmonics of the first overtone. The harmonics, except $2v_1$, are not detected. Such pattern is very different from modulation patterns observed in RRc stars \citep{moje_bl}. Beating scenario is more realistic for both stars: we observe two close frequencies, their harmonics and linear combinations of the two signals. Period ratio of two signals is 0.9788 and 0.9904 for OGLE-BLG-RRLYR-11754 and OGLE-BLG-RRLYR-15059, accordingly, which suggest that at least one signal cannot correspond to the radial mode. In the middle panels of Figs. \ref{Fig.11754_lc} and \ref{Fig.15059_lc} we plotted disentangled light curve for the dominant periodicity for OGLE-BLG-RRLYR-11754 and OGLE-BLG-RRLYR-15059, respectively. In the bottom panels we present light curves corresponding to the additional lower amplitude signal. In both stars, and in V37 \citep[see fig. 12 in][]{ngc6362}, light curves corresponding to the dominant periodicity are very symmetric, whereas light curves corresponding to the lower amplitude signal are more non-sinusoidal. We calculated Fourier parameters for light curves corresponding to the dominant and additional periodicity of both stars. Their values are not typical for RRc stars. Especially phase differences are lower than for RRc stars. The nature of all three stars remains a puzzle. We refer the reader to section 3.7 in \cite{ngc6362} for a more detailed discussion.

\begin{table}
\centering
\caption{Two interpretations for OGLE-BLG-RRLYR-11754. First three columns list frequencies, amplitudes and phases of the significant peaks detected in the frequency spectrum of OGLE-BLG-RRLYR-11754. The following two columns provide two possible interpretations: `modulation' and `beating' scenarios.}
\label{tab:11754}
\begin{tabular}{r@{.}lllll}
\multicolumn{2}{c}{$\nu$\thinspace (c\,d$^{-1}$)} &  $A$\thinspace (mag)     & $\phi$\thinspace (rad)   & modulation?           & beating?\\
\hline
0&07989431 & 0.0049 & 0.90 & $\nu_{\rm m}$          & $\nu_{\rm 1}-\nu_x$ \\ 
3&68319790 & 0.0650 & 1.94 & $\nu_1-\nu_{\rm m}$    & $\nu_x$ \\ 
3&76309221 & 0.066343 & 1.26 & $\nu_1$               & $\nu_1$ \\ 
7&36639581 & 0.009335 & 0.31 & $2\nu_1-2\nu_{\rm m}$    & $2\nu_x$ \\
7&44629012 & 0.004961 & 5.70 & $2\nu_1-\nu_{\rm m}$    & $\nu_1+\nu_{\rm x}$ \\ 
7&52618443 & 0.004394 & 4.68 & $2\nu_1$   & $2\nu_1$ \\ 
11&04959371 & 0.004867 & 4.84 & $3\nu_1-3\nu_{\rm m}$              & $3\nu_x$ \\ 
14&73279162 & 0.002674 & 3.21 & $4\nu_1-4\nu_{\rm m}$   & $4\nu_x$ \\
\hline
\end{tabular}
\end{table}

\begin{table}
\centering
\caption{The same as Tab.~\ref{tab:11754}, but for OGLE-BLG-RRLYR-15059.}
\label{tab:15059}
\begin{tabular}{r@{.}lllll}
\multicolumn{2}{c}{$\nu$\thinspace (c\,d$^{-1}$)} &  $A$\thinspace (mag)     & $\phi$\thinspace (rad)   & modulation?           & beating?\\
\hline
 3&76105696 & 0.074649 & 4.98 & $\nu_1$ & $\nu_1$ \\
 7&52211392 & 0.005627 & 0.24 & $2\nu_1$ & $2\nu_1$ \\
 3&72483401 & 0.049168 & 3.97 & $\nu_1-\nu_{\rm m}$ & $\nu_{\rm x}$ \\
 7&48589097 & 0.006323 & 4.82 & $2\nu_1-\nu_{\rm m}$ & $\nu_1+\nu_{\rm x}$ \\
 0&03622295 & 0.003700 & 2.78 & $\nu_{\rm m}$          & $\nu_{\rm 1}-\nu_x$ \\ 
 7&44966801 & 0.011156 & 4.48 & $2\nu_1-2\nu_{\rm m}$    & $2\nu_x$ \\
11&17450202 & 0.004099 & 4.20 & $3\nu_1-3\nu_{\rm m}$              & $3\nu_x$ \\
 14&89933602 & 0.003803 & 4.94 & $4\nu_1-4\nu_{\rm m}$   & $4\nu_x$ \\

\hline
\end{tabular}
\end{table}

\section{Summary and conclusions}\label{sec.summary}

We analyzed the complete sample of RRc and RRd stars from the Galactic bulge based on the OGLE-IV data. Summary of the results is as follows:
\begin{itemize}
\item We have detected 960 $RR_{0.61}$ stars, i.e. stars with additional, low-amplitude variability and period ratios, $P_x/P_{\rm 1O}$, in the $0.6-0.64$ range. The incidence rate of this phenomenon, considering the whole OGLE sample is thus 8.3 per cent. The real incidence rate is much higher however. Our previous analysis of selected OGLE-IV fields with the highest observing cadence and hence of relatively low noise points towards incidence rate of 27 per cent \citep{moje_o4_061}. Space photometry, on the other hand, indicates that $RR_{0.61}$ must be common, as nearly all RRc/RRd stars observed from space belong to $RR_{0.61}$ class \citep{061_k2, 068kepler}. 
\item In the Petersen diagram $RR_{0.61}$ stars form three well defined sequences. Bottom sequence is the most populated and is centered at $P_x/P_{\rm 1O}=0.613$. Middle sequence contains lowest number of stars and is centered at $P_x/P_{\rm 1O}=0.622$. The top sequence is not sharply defined. It is centered at $P_x/P_{\rm 1O}=0.631$.
\item In 114 $RR_{0.61}$ stars, in addition to low amplitude signals at $f_x$, we detect signals centered at $0.5f_x$. Of these stars 111 are RRc and 3 are RRd. Signals at $0.5f_x$ correspond to the top sequence in the Petersen diagram for the majority of stars. This finding supports explanation for additional signals proposed by \cite{dziembowski}. In this explanation, signals observed at $0.5f_x$ are non-radial modes of $\ell=8$ (top sequence) and $\ell=9$ (bottom sequence). Due to geometric cancellation, modes of $\ell=8$ should be easier to detect and this is what we observe. In this theory, signals detected at $f_x$ are harmonics of non-radial modes that do not suffer from geometric cancellation.
\item In seven stars we detect additional signals that would fit the expected position of $\ell=8$ non-radial mode. Harmonic is not observed in these stars however.
\item In 134 $RR_{0.61}$ stars we observe signals corresponding to more than one sequence on the Petersen diagram. 101 stars have signals corresponding to two sequences and 33 stars have signals corresponding to three sequences. According to the model proposed by \cite{dziembowski}, the middle sequence in the Petersen diagram arises due to linear combination frequency of the $\ell=8$ and $\ell=9$ modes. For stars in which we detect three signals simultaneously, we confirm that the middle signal is indeed a combination - another point in favor of the Dziembowski model.
\item We have detected 147 $RR_{0.68}$ stars, i.e. stars with the additional low-amplitude variability and period ratios $P_{\rm 1O}/P_x$ close to 0.686. The incidence rate is thus 1.3 per cent. 128 stars are new detections. We note that we still lack the interpretation for this mysterious group in which additional signal has period longer that the period of the radial fundamental mode (not detected in these stars).

\item We found two RRc stars, similar to V37 from the globular cluster NGC~6362 \citep{ngc6362}, in which we observe beating of two close frequencies. Dominant frequency has sinusoidal light-curve and the additional period corresponds to more non-linear light curve characteristic for pulsations in the fundamental mode. Both light curves differ from typical for RRc stars. As in the case of V37, there is no satisfactory explanation for these stars yet.
\end{itemize}

\section*{Acknowledgements}

This research is supported by the Polish Ministry of Science and Higher Education under grant 0192/DIA/2016/45 within the Diamond Grant Programme for years 2016-2020 (HN) and by the National Science Center, Poland, grant agreement DEC-2015/17/B/ST9/03421 (RS).




\bibliographystyle{mnras}
\bibliography{biblio} 




\appendix

\section{Data tables}

\begin{onecolumn}
\begin{longtable}{cllllll}
\caption{Properties of RR$_{0.61}$ stars. Consecutive columns provide name of a star, its first-overtone period, period of the additional signal, ratio between the two periods, amplitude of the first overtone and ratio of amplitude of the additional signal to the amplitude of the first overtone mode. Last column contains remarks on individual stars: `comb.' - there is a combination between the first overtone and the additional signal, `bl' - Blazhko effect is present in the star, `cand.' - additional signal is weak (around $S/N=4$) or its daily aliases are higher, $0.5f_x$ - subharmonic of the additional signal is detected.}\\
ID & $P_{\rm 1O}$\thinspace [d] &  $P_{\rm X}$\thinspace [d] & $P_{\rm 1O}/P_{\rm X}$  & $A_{\rm 1O}$\thinspace [mag]   &  $A_{\rm X}/A_{\rm 1O}$   & Remarks \\
\hline
\endfirsthead
{\tablename\ \thetable\ -- \textit{Continued from previous page}} \\
ID & $P_{\rm 1O}$\thinspace [d] &  $P_{\rm X}$\thinspace [d] & $P_{\rm 1O}/P_{\rm X}$  & $A_{\rm 1O}$\thinspace [mag]   &  $A_{\rm X}/A_{\rm 1O}$   & Remarks \\
\hline
\endhead
\hline \multicolumn{6}{r}{\textit{Continued on next page}} \\
\endfoot
\hline
\endlastfoot
OGLE-BLG-RRLYR-00368 & 0.27303168(3) & 0.1674370(8) & 0.61325 & 0.1013(6) & 0.031 & cand. \\ 
OGLE-BLG-RRLYR-00710 & 0.28749643(7) & 0.180509(1) & 0.62787 & 0.1420(7) & 0.021 &  \\ 
OGLE-BLG-RRLYR-00815 & 0.3117463(3) & 0.192034(1) & 0.61599 & 0.120(2) & 0.041 &  \\ 
OGLE-BLG-RRLYR-01014 & 0.30330353(6) & 0.186214(1) & 0.61395 & 0.1226(4) & 0.022 &  \\ 
OGLE-BLG-RRLYR-01067 & 0.28294809(2) & 0.1736623(8) & 0.61376 & 0.1426(2) & 0.013 &  \\ 
OGLE-BLG-RRLYR-01088 & 0.29471870(5) & 0.1811678(9) & 0.61471 & 0.1319(4) & 0.021 &  \\ 
OGLE-BLG-RRLYR-01097 & 0.2923084(4) & 0.179451(1) & 0.61391 & 0.1357(5) & 0.026 &  \\ 
OGLE-BLG-RRLYR-01099 & 0.25533780(4) & 0.1564733(2) & 0.61281 & 0.1097(3) & 0.036 &  \\
\end{longtable}
\end{onecolumn}

\begin{onecolumn}
\begin{longtable}{cllllll}
\caption{Properties of RR$_{0.68}$ stars. Consecutive columns provide name of a star, its first-overtone period, period of the additional signal, ratio between the two periods, amplitude of the first overtone and ratio of amplitude of the additional signal to the amplitude of the first overtone mode. Last column contains remarks on individual stars: `comb.' - there is a combination between the first overtone and the additional signal, `bl' - Blazhko effect is present in the star.}\\

ID & $P_{\rm 1O}$\thinspace [d] &  $P_{\rm X}$\thinspace [d] & $P_{\rm X}/P_{\rm 1O}$  & $A_{\rm 1O}$\thinspace [mag]   &  $A_{\rm X}/A_{\rm 1O}$   & Remarks \\
\hline
\endfirsthead
{\tablename\ \thetable\ -- \textit{Continued from previous page}} \\
ID & $P_{\rm 1O}$\thinspace [d] &  $P_{\rm X}$\thinspace [d] & $P_{\rm 1O}/P_{\rm X}$  & $A_{\rm 1O}$\thinspace [mag]   &  $A_{\rm X}/A_{\rm 1O}$   & Remarks \\
\hline
\endhead
\hline \multicolumn{6}{r}{\textit{Continued on next page}} \\
\endfoot
\hline
\endlastfoot
OGLE-BLG-RRLYR-01064 & 0.31486155(4) & 0.458607(6) & 0.68656 & 0.1239(3) & 0.016 &  \\ 
OGLE-BLG-RRLYR-01152 & 0.34759513(10) & 0.507535(4) & 0.68487 & 0.1084(5) & 0.04 & comb. \\ 
OGLE-BLG-RRLYR-01408 & 0.3853534(3) & 0.56176(1) & 0.68598 & 0.112(1) & 0.04 &  \\ 
OGLE-BLG-RRLYR-01436 & 0.3242810(5) & 0.47942(2) & 0.6764 & 0.120(3) & 0.04 &  \\ 
OGLE-BLG-RRLYR-01808 & 0.3161546(2) & 0.460225(5) & 0.68696 & 0.0937(8) & 0.063 &  \\ 
OGLE-BLG-RRLYR-03053 & 0.2898407(1) & 0.423478(6) & 0.68443 & 0.1273(9) & 0.042 & bl \\ 
OGLE-BLG-RRLYR-03058 & 0.3862909(2) & 0.56395(1) & 0.68497 & 0.1342(9) & 0.019 & bl \\ 
OGLE-BLG-RRLYR-03581 & 0.3763837(3) & 0.549241(5) & 0.68528 & 0.0964(9) & 0.059 &  \\
\end{longtable}
\end{onecolumn}


\bsp	
\label{lastpage}
\end{document}